\documentclass[twocolumn, prl, aps, superscriptaddress,shortbibliography,showpacs,amsmath,amssymb,floatfix]{revtex4-1}
\usepackage{graphicx}
\usepackage{amssymb}
\usepackage{amsmath}
\usepackage{epsfig}
\usepackage{color}
\usepackage{tabu}
\usepackage{mathtools}
\usepackage[colorlinks,linkcolor=blue,anchorcolor=blue,citecolor=blue,urlcolor=blue]{hyperref}
\usepackage{physics}
\usepackage{float}
\usepackage{diagbox}
\DeclareMathOperator{\Hc}{H.c.}

\begin{document}

\title{Bose-Einstein condensate in Bloch bands with off-diagonal periodic potential}
\author{Yue-Xin Huang}
\affiliation{CAS Key Laboratory of Quantum Information, University of Science and Technology of China, Hefei, 230026, China}
\author{Wei Feng Zhuang}
\affiliation{CAS Key Laboratory of Quantum Information, University of Science and Technology of China, Hefei, 230026, China}
\author{Xiang-Fa Zhou}
\affiliation{CAS Key Laboratory of Quantum Information, University of Science and Technology of China, Hefei, 230026, China}
\affiliation{Synergetic Innovation Center of Quantum Information and Quantum Physics, University of Science and Technology of China, Hefei, Anhui 230026, China}
\affiliation{CAS Center For Excellence in Quantum Information and Quantum Physics}
\author{Han Pu}
\affiliation{Department of Physics and Astronomy, Rice University, Houston, Texas 77005, USA}
\author{Guang-Can Guo}
\affiliation{CAS Key Laboratory of Quantum Information, University of Science and Technology of China, Hefei, 230026, China}
\affiliation{Synergetic Innovation Center of Quantum Information and Quantum Physics, University of Science and Technology of China, Hefei, Anhui 230026, China}
\affiliation{CAS Center For Excellence in Quantum Information and Quantum Physics}
\author{Ming Gong}
\email{gongm@ustc.edu.cn}
\affiliation{CAS Key Laboratory of Quantum Information, University of Science and Technology of China, Hefei, 230026, China}
\affiliation{Synergetic Innovation Center of Quantum Information and Quantum Physics, University of Science and Technology of China, Hefei, Anhui 230026, China}
\affiliation{CAS Center For Excellence in Quantum Information and Quantum Physics}
\date{\today }

\begin{abstract}
 We report the Bose-Einstein condensate (BEC) in the Bloch bands with off-diagonal periodic potential (ODPP), which simultaneously plays the role of spin-orbit coupling (SOC) and Zeeman field. This model can be realized using two independent Raman couplings in the same three level system, in which 
	the time-reversal symmetry ensures the energy degeneracy between the two states with opposite momenta. 
    We find that these two Raman couplings can be used to tune the spin polarization in momentum space, thus greatly modifies the effective scatterings 
	over the Bloch bands. We observe a transition from the  Bloch plane wave phase with condensate at one wave vector to the Bloch stripe phase with 
	condensates at the two Bloch states with opposite wave vectors. These two phases will exhibit totally different spin textures and density modulations in 
	real space, which are totally different from that in free space. In momentum space multiple peaks differ by some reciprocal lattice vectors can be 
	observed, reflecting the periodic structure of the ODPP. A three-band effective model is proposed to understand these observations. 
	This system can provide a new platform in investigating of various physics, such as collective excitations, polaron and topological superlfuids, over 
	the Bloch bands.
\end{abstract}

\maketitle

Spin-orbit coupling (SOC), which couples the spin and momentum degrees of freedoms, plays an important role in many
important concepts in condensed matter physics \cite{hasan_colloquium_2010,qi_topological_2011, sinova2015spin, xiao2010berry}. In recent years, this interaction has been widely sought in
ultracold atoms \cite{galitski_spin-orbit_2013,stuhl_visualizing_2015,zhai_degenerate_2015,fu_production_2013,wall_synthetic_2016,kolkowitz_spinorbit-coupled_2017}.
In Bose gases, it could be used to realize Bose-Einstein condensate (BEC) \cite{anderson_observation_1995,bradley_evidence_1995}
with finite momentum, which belongs to either plane wave phase or stripe phase \cite{wang_spin-orbit_2010,li_quantum_2012,li_stripe_2017}, depending
strongly on the inter-particle and intra-particle interaction strengths. The spin dipole in this system will also exhibits some exotic behavior in quench
dynamics \cite{li_spin_2018}. In Fermi gases, it can be used for the realization of topological superfluids
\cite{qi_topological_2011,fu_superconducting_2008,cooper_stable_2009,oreg_helical_2010,lutchyn_majorana_2010,zhou_topological_2011,mourik_signatures_2012}
and associated Majorana zero modes \cite{kitaev_unpaired_2001,tewari_quantum_2007,sun_majorana_2016}, due to the effective $p$-wave pairing at
the Fermi surface. By carefully engineering the interaction, this system may also be used to create different types of gapless phases \cite{hsieh_topological_2008,
pesin_mott_2010, gong_bcs-bec_2011, huang2015large}. To date both one \cite{lin_spin-orbit-coupled_2011} and two dimensional SOC have been realized with Bosons \cite{wu_realization_2016}
and Fermions \cite{huang_experimental_2016}. In these experiments, SOC is realized by
Raman coupling \cite{zhu_spin_2006,spielman_raman_2009,liu_effect_2009,anderson_synthetic_2012,zhai_degenerate_2015},
which can be brought into Rashba or Dresselhaus SOC form by a unitary transformation.
These progresses opened a new avenue for searching of exotic phases in degenerate gases \cite{lin_spin-orbit-coupled_2011, zhang_collective_2012, wang_spin-orbit_2012,hamner_dicke-type_2014,
ji_experimental_2014,olson_tunable_2014,li_spin-orbit_2016}.

\begin{figure}[h]
    \centering
    \includegraphics[width=0.38\textwidth]{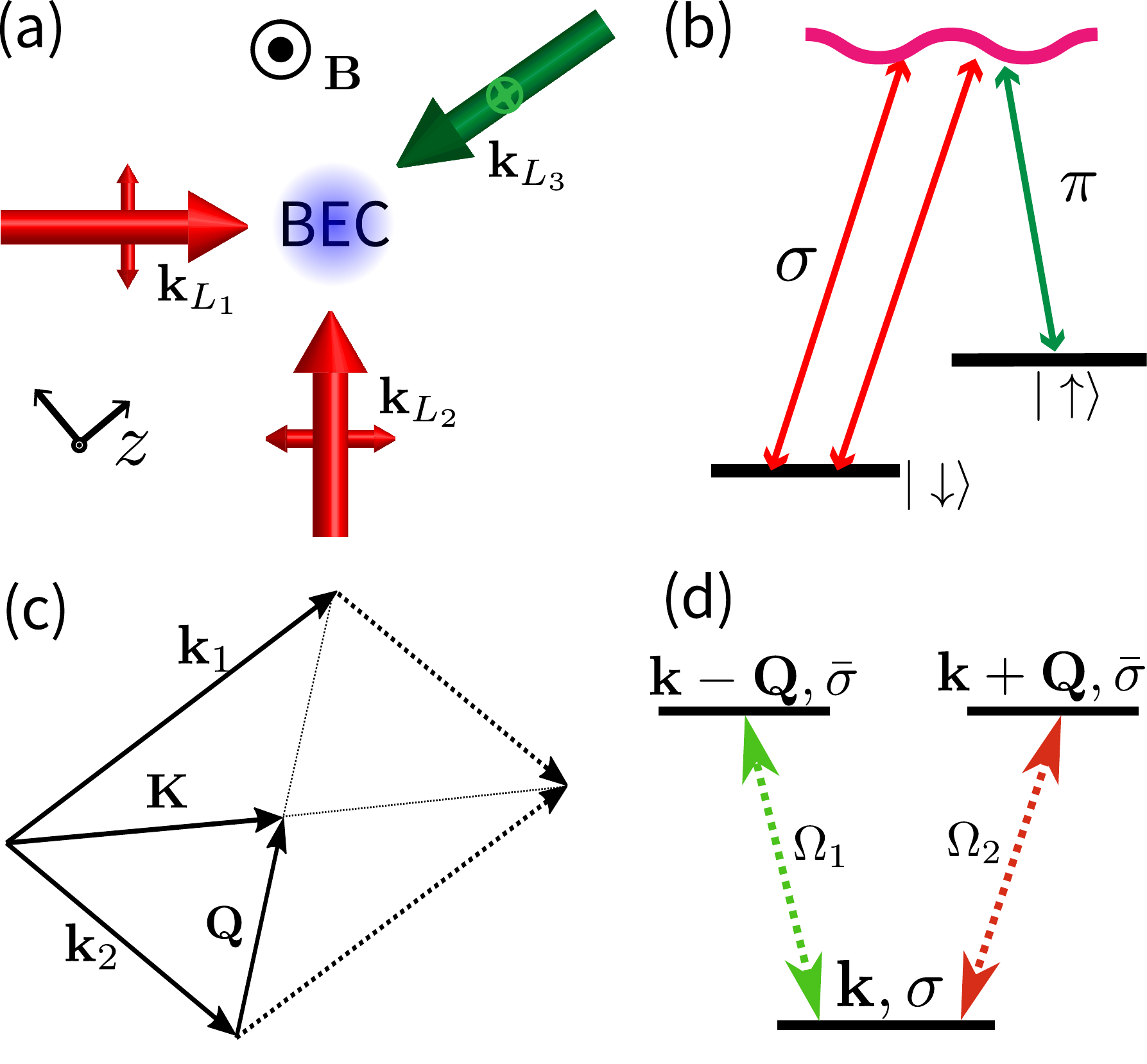}
	\caption{(a) Setup for the ODPP. The laser $L_1$ with frequency $\omega_1$ is linearly polarized along $z$ direction, and the
    two lasers $L_{2/3}$ with identical frequency $\omega_2$ from the same source are orthogonally polarized in the $x$-$y$ plane.
    (b) These three laser beams form two independent Raman coupling in the same three level system.
    (c) Momentum transfer in these two Raman couplings with $\vb k_i\equiv \vb k_{L_i}-\vb k_{L_3}$ for $i = 1$, 2.
    (d) The plane wave ${\bf k}$ with spin $\sigma$ is coupled the two other plane wave vectors ${\bf k} \pm {\bf Q}$ with opposite spin 
	$\bar{\sigma}$ by $\Omega_1$ and $\Omega_2$, respectively.}
    \label{fig-fig1}
\end{figure}

\begin{figure}[h]
    \centering
    \includegraphics[width=0.47\textwidth]{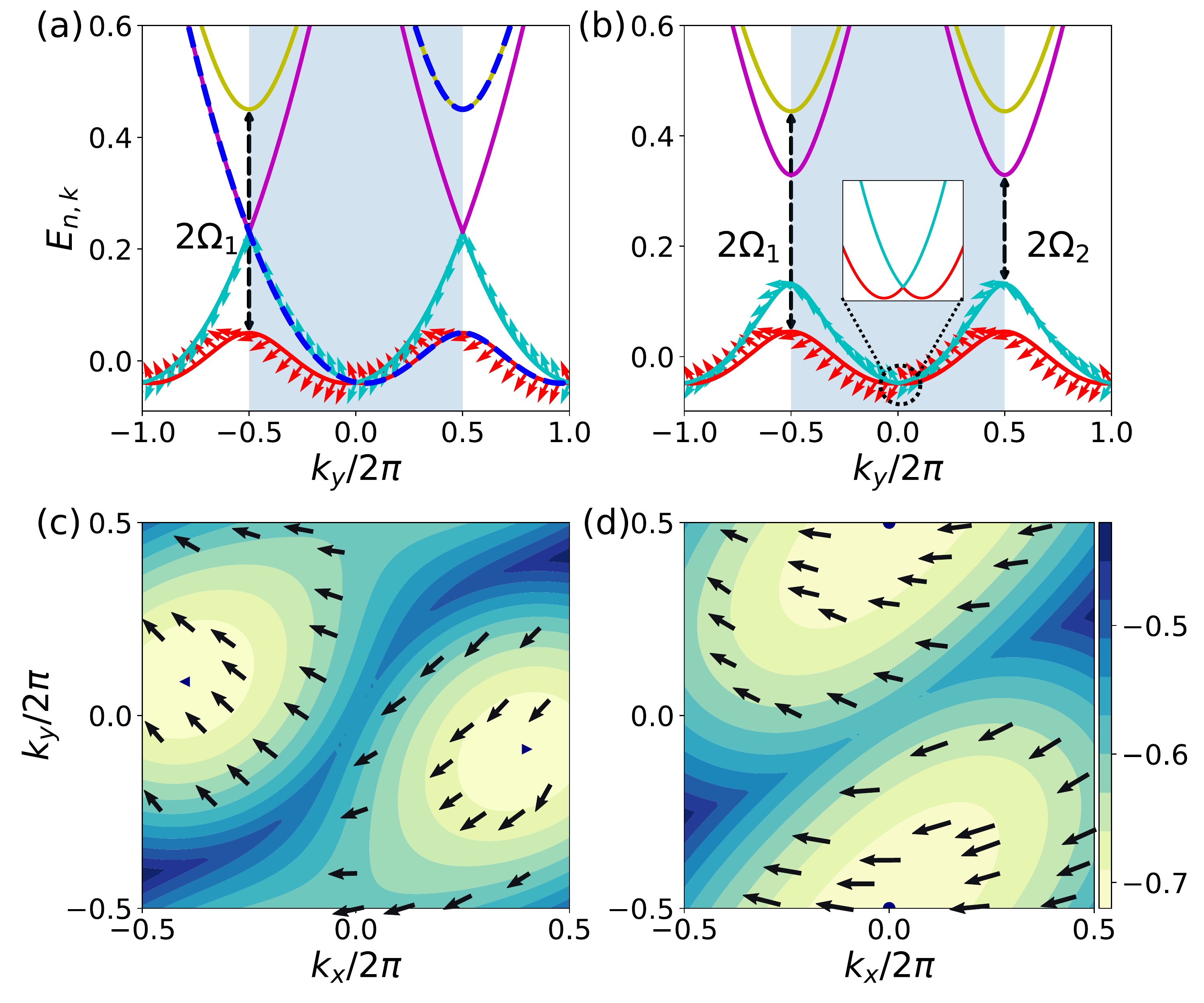}
    \caption{(a) Energy spectrum with $\Omega_1=0.2$, $\Omega_2=0$ when $k_x=0$.
        The blue dashed lines are obtained with $\tilde{\mathcal H}$, in which $k_y$
        has been shift to match the bands obtained using Bloch theorem.
        (b) Spectrum with $\Omega_1=0.2$, $\Omega_2=0$ when $k_x=0$. The inset shows the detail of the double minima.
        (c) Double minima spectrum in two dimension when $\Omega_1=0.6$ and $\Omega_2=0.2$. 
        (d) Single minimum at $(0, \pi)$ with $\Omega_1=1.2$ and $\Omega_2=0.2$. The arrows in (c) and (d) represent 
		the spin polarization in $\sigma_x$-$\sigma_z$ space in momentum spacing, indicating of spin-momentum locking from two 
		dimensional SOC.}
    \label{fig-fig2}
\end{figure}

In this work, we consider the fate of a BEC in the Bloch bands with off-diagonal periodic potential (ODPP), which plays the role of SOC and Zeeman field simultaneously.
This potential can be realized using three laser beams coupled to the same three level system (in $\Lambda$ configuration),
which forms two sets of Raman couplings. In this model the position of the local energy minima and their corresponding
spin textures can be tuned by the two Raman couplings, which in turn greatly influence the scattering of the Bloch states.
This model exhibits rich phase structures in both the single particle band structure and the interacting condensates.
Especially, we find a transition from the spin-imbalanced Bloch plane wave (BPW) phase condensed at one Bloch wave vector ${\bf k}_0$ to the
spin-balanced  Bloch stripe (BST) phase condensed at two vectors $\pm {\bf k}_0$. These two phases exhibit multiple peaks
differ by some reciprocal lattice vectors in momentum space. Meanwhile, they will also exhibit some intriguing spin textures and 
density modulations in real space. Thus these two phases will have features totally different from that in free space.
This new platform also serve as an interesting model for exploring various new physics in the Bloch bands.

{\it Model and Hamiltonian}. We consider a $^{87}$Rb BEC in a weak trap (see Fig. \ref{fig-fig1} (a)). The magnetic field along $\vu z$ 
direction sets the spin quantization axis. Three laser beams are used to couple the ground
state manifold $\ket{F=1,m_F}$ in 5$^2\mathrm{S}_{1/2}$ to the excited state manifold in 5$^2\mathrm{P}_{3/2}$ to construct two sets of Raman couplings.
Here laser beam $L_3$ is linearly polarized along $\vu z$ direction ($\pi$ transition) and $L_1$ and $L_2$ from the same laser
source with identical frequency are polarized in the $x$-$y$ plane ($\sigma$ transition). We assume that the polarization of the $L_1$ and $L_2$ beams are mutually orthogonal to avoid interference. These two couplings are accompanied by momentum
transfer $\tilde{\vb k}_i\equiv \vb k_{L_i}-\vb k_{L_3}$ for $i=1, 2$. By labeling $\ket{\downarrow}=\ket{1,-1}$ and $\ket{\uparrow}=\ket{1,0}$,
we have Hamiltonian,
\begin{eqnarray}
    H=\int\dd \vb r\Psi^\dagger(\vb r)\left[ \mathcal{H}(\vb r)+\mathcal{V}_\mathrm{trap}(\vb r) \right]\Psi(\vb r).
    \label{eq-h}
\end{eqnarray}
By eliminating the excited bands in the large detuning limit, we obtain the following single particle Hamiltonian,
\begin{eqnarray}
    \mathcal{H}(\vb r)=\mqty(\frac{\tilde{\vb p}^2}{2m}+\tilde\delta & \sum_{j=1,2}\tilde \Omega_j e^{i\tilde{\vb k}_j\cdot \tilde{\vb r} } \\
    \sum_{j=1,2}\tilde \Omega_j^*e^{-i\tilde{\vb k}_j\cdot \tilde{\vb r}} & \frac{\tilde{\vb p}^2}{2m}-\tilde\delta),
    \label{eq-matcalh}
\end{eqnarray}
under basis $\Psi(\vb r)=(\psi_\uparrow(\vb r), \psi_\downarrow(\vb r))^T$, where $\mathcal{V}_\mathrm{trap}(\vb r)$ is the
harmonic trap potential, $m$ is the mass and $\tilde{\vb p}$ is the momentum operator, $\tilde\Omega_j$ is the resonant
Raman coupling strength, and $\tilde \delta$ is the detuning from Raman resonance.  Note that the phases carried by $\tilde{\Omega}_{1,2}$ are always fixed as $L_1$ 
and $L_2$ originate from the same laser source, making our model immune from random phase fluctuation. When $\tilde \Omega_2$ (or $\tilde\Omega_1$) equals
to zero, Eq. \ref{eq-matcalh} is reduced to the well-known one dimensional model with SOC by performing a unitary transformation
\cite{zhai_degenerate_2015,galitski_spinorbit_2013,sinova_spin_2015}. In the presence of both Raman couplings, the phase 
carried by ODPP can no longer be gauged out.

In following, we rescale the energy and momentum in units of recoil energy $E_r=\hbar^2k_r^2/2m$ and recoil momentum $k_r$ \cite{aspect_laser_1988}, 
then we can define $\tilde\delta=\delta E_r$, ${\tilde\Omega}_j=\Omega_j E_r$, $\tilde{\vb k}_i= \vb k_ik_r$,
$\tilde{\vb K}={\vb K} k_r$ and $\tilde{\vb Q}={\vb Q} k_r$, where $\vb K=\frac{\vb k_1+\vb k_2}{2}$ and $\vb Q=\frac{\vb k_1-\vb k_2}{2}$ (see 
Fig. \ref{fig-fig1} (c)). After a unitary transformation \cite{galitski_spinorbit_2013}, we have
\begin{eqnarray}
    \mathcal{H}(\vb r)=
    \mqty( {(\vb p+\vb K/2)^2}+\delta & \Omega_1 e^{i\vb Q\cdot \vb r}+\Omega_2 e^{-i\vb Q\cdot \vb r}\\
    \Omega_1^* e^{-i\vb Q\cdot \vb r}+\Omega_2^* e^{i\vb Q\cdot \vb r} & {(\vb p-\vb K/2)^2}-\delta).
    \label{eq-h2}
\end{eqnarray}
The physical meaning now becomes clear, the global plane carried by vector ${\bf K}$ plays the role of one dimensional SOC, and the 
ODPP, which represents a helical magnetic field with period determined by $2\pi/Q$ in real space, plays the role of Zeeman field 
locally. However, it is more complicated because this helical magnetic field is coupled to momentum, hence behaving as a source of SOC itself. The two dimensional nature of the SOC from the spin-momentum locking effect is shown in Fig.~\ref{fig-fig2}(c) and (d). The relative phase between the two Raman couplings can be gauged out by a position shift, thus is 
unimportant. This model is invariant under the anti-unitary transformation $\Theta = \sigma_x \mathcal K$, where $\mathcal{K}$ 
is the complex conjugate operator. In the following, for simplicity,
we only report the case of $\abs{\vb k_1}=\abs{\vb k_2}$, then $\vb K$ is perpendicular to $\vb Q$. The case with two non-perpendicular 
vectors will be discussed elsewhere. In our simulation, we let $\vb K= K\vu x$ and $\vb Q=Q\vu y$, then the diagonal term represents 
the usual one dimensional SOC as $k_x\sigma_z$ along $x$ direction.

\begin{figure}[h]
    \centering
    \includegraphics[width=0.42\textwidth]{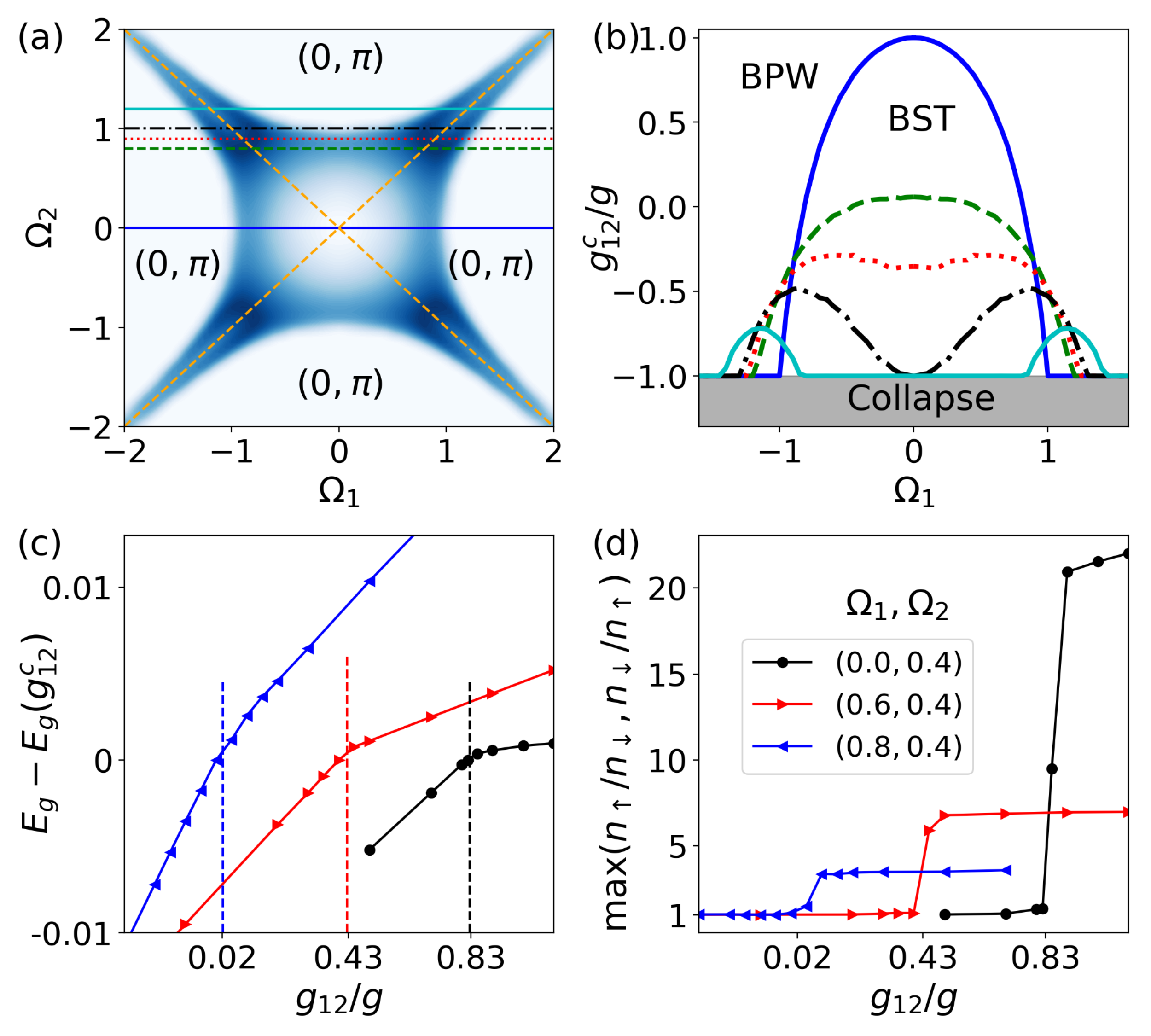}
    \caption{(a) Single particle phase diagram in the parameter space $\Omega_1$ and $\Omega_2$. The 
	color represents the distance between the two minima, thus the white color represents one local minimum. 
	(b) The phase boundary between BPW ($g_{1} > g_{12}^c$) and BST ($g_{12} < g^c_{12}$) influenced by the 
	two Raman couplings for the three horizon lines in (a), with $\Omega_2 = (0,  0.8, 0.9, 1.0, 1.2)$. 
    The system collapses when $g_{12}/g<-1$. 
	(c) and (d) Ground state energy and spin polarization across the phase boundaries for three typical 
	parameters.}
    \label{fig-fig3}
\end{figure}

Due to the nature of ODPP, we write the wave function using plane wave basis as \cite{kittel_introduction_2005}
\begin{eqnarray}
	\psi_{s,\vb k}(\vb r)=\sum_{\vb G}\phi_{s,\vb k,\vb G} e^{i(\vb k+\vb G)\cdot \vb r},
    \label{eq-blochwf}
\end{eqnarray}
by the Bloch theorem. The quasi momentum $\vb k$ is a good quantum number restricted to the first Brillouin zone (BZ), so $k_y\in[-Q/2,Q/2]$;
and $\vb G=n\vb Q$ ($n\in \mathbb{Z}$) are the reciprocal lattice vectors. 
In this basis, the Hamiltonian can be written as $H_{\vb k}=\sum_{\vb G}\mathcal{H}_{\vb k}({\bf G})$, where
\begin{eqnarray}
	&& \mathcal{H}_{\vb k}(\vb G)= \sum_{s=\uparrow,\downarrow}\phi_{s,\vb k, \vb G}^\dagger\left[ \left( \vb k+\vb G+c_s\vb K/2 \right)^2 \right]\phi_{s,\vb k,\vb G}
    \nonumber\\
	&&+\Big[ \Omega_1\phi_{\uparrow,\vb k, \vb G}^\dagger \phi_{\downarrow,\vb k,\vb G-\vb Q}+\Omega_2\phi_{\uparrow,\vb k,\vb G}^\dagger \phi_{\downarrow,\vb k, \vb G+\vb Q}+\Hc \Big],
    \label{eq-hamiltonian_k}
\end{eqnarray}
with $c_\uparrow=1$ and $c_\downarrow=-1$. We see that the $\Omega_1$ field couples $\phi_{\uparrow,\vb k, \vb G}$
to $\phi_{\downarrow,\vb k, \vb G-\vb Q}$, while the $\Omega_2$ field couples $\phi_{\uparrow,\vb k, \vb G}$ to $\phi_{\downarrow,\vb k, \vb G + \vb Q}$ 
(see illustration in Fig. \ref{fig-fig1} (d)). In momentum space, one can define a TR operator $\Theta_\text{tr} =\mathbb{I}_{2n_\text{c}+1}\otimes \Theta$, 
with $\Theta_\text{tr} {H}_{\vb k}\Theta_\text{tr}^{-1}= {H}_{-\vb k}$ \cite{ando_topological_2013}. Moreover the Hamiltonian 
can be made real if all $\Omega_i \in\mathbb{R}$ using the symmetry $\mathcal{K}$. 

\begin{figure}[h]
    \centering
	\includegraphics[width=0.48\textwidth]{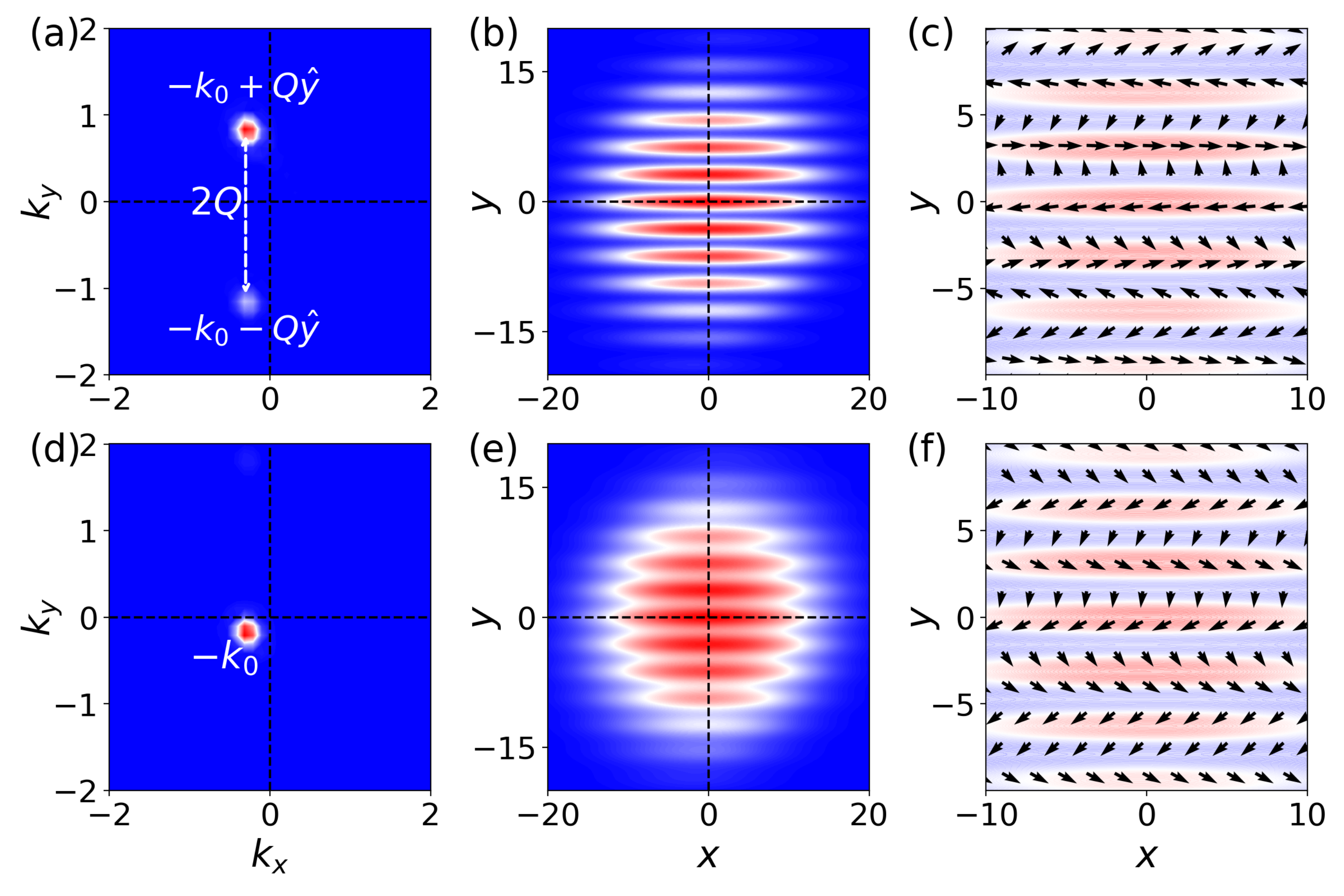}
	\caption{Properties of the BPW phase. (a) and (d) Wave function in momentum space for spin up and spin down, respectively. (b) and 
	(e) The corresponding wave functions for these two components in real space. Spin polarizarion in $\sigma_x - \sigma_y$ space and    $\sigma_x - \sigma_z$ space 
	are shown in (c) and (f). These results are obtained from GPE simulation with parameters: $\Omega_1 = 0.4$, $\Omega_2 = 0.8$, $g_{12}/g = 0.1$ and 
	$\vb k_0\approx(0.30,0.15)$.}
	\label{fig-fig4}
\end{figure}

{\it Single-particle phase diagram}. We first consider the single-particle spectra, as shown in Fig. \ref{fig-fig2}. The case with $\Omega_2 = 0$ 
can be obtained exactly via a unitary transformation, which yields $\tilde{\mathcal{H}}=\vb k^2+\sigma_z\vb k\cdot \vb k_1+\Omega_1\sigma_x$
and spectra $\epsilon_{{\bf k} \pm} = \vb k^2\pm \sqrt{(\vb k\cdot \vb k_1)^2+\Omega_1^2}$ (see the blue dashed lines in Fig. \ref{fig-fig2} (a)). 
However, by Bloch theorem, the dispersion in free space should be folded into the first BZ, in which the Raman coupling $\Omega_1$ will open 
an energy gap between the ground state and third excited band (see Fig. \ref{fig-fig2}(a)). Other degeneracies at the high symmetry 
points come from level crossing during folding of bands. The folded spectra form the first and second excited bands. 

Next we switch on the second Raman coupling $\Omega_2\ne 0$, and the corresponding Bloch bands are shown in Fig.~\ref{fig-fig2}(b). 
We see that this additional coupling can open an energy gap between the folded spectra at $k_y= \pm {1\over 2}$. 
The interplay between these two Raman couplings can greatly influence the spin texture in momentum space, similar to the situation with two dimensional SOC \cite{dresselhaus_spin-orbit_1955, bihlmayer_focus_2015}. However, the real nature of $\mathcal{H}_{\bf k}$
prohibits the existence of geometry phase in a closed loop. Two typical examples for the lowest
band $E_{1,{\bf k}}$ from the Bloch wave functions are presented in Fig. \ref{fig-fig2}(c) - (d), from which one can see that the position of the ground state
minima can be controlled in the whole BZ by tuning the two Raman coupling strengths. Moreover, the TR symmetry $\Theta_\text{tr}$ ensures 
$E_{n,{\bf k}} = E_{n,-{\bf k}}$.

The phase diagram of the single-particle Hamiltonian is characterized by the position of the energy minima, and is presented in 
Fig.~\ref{fig-fig3}(a), which exhibits a star structure. To understand this diagram, let us consider the limit that $|\Omega_1|\gg |\Omega_2|$,
then by ignoring $\Omega_2$, the Hamiltonian can be written as $k^2 + (Q k_y + K k_x) \sigma_z + \Omega_1 \sigma_x$,
which exhibits two local minima when $\Omega_1$ is small, and one minimum at ${\bf k} = 0$ when $|\Omega_1|$ is much larger than $\abs{Q \vu y + K \vu x}$. 
Noticed that the unitary transformation has introduced a momentum shift, thus the single minimum is shifted to ${\bf k} = (0, \pi)$. 
When $\Omega_1= \pm \Omega_2$, which corresponds to the diagonal and off-diagonal axes in Fig.~\ref{fig-fig3}(a) denoted by dashed lines, 
it will always exhibit two local minima even when $\Omega_1$ becomes large. In this case, these two minima will never merge to a single mimimum. 
Thus by tuning these two Raman couplings, one can not only engineer the spin polarization, but also the position of the
ground state minima, which can influence the fate of the BEC over these bands.

{\it BEC over the Bloch bands and phase diagram}. With these features, we naturally ask the question: What will happen to the condensate in these Bloch bands? 
From the viewpoint of plane wave basis, this kind of condensate occupies multiple momenta simultaneously. In the weak interacting limit, one expect 
the atoms to be condensed at the ground state(s) of the Bloch bands. We consider the following interaction \cite{inouye_observation_1998},
\begin{eqnarray}
	\mathcal{V}_\mathrm{I}=\int\dd \vb r\left[ g\left( n_\uparrow^2(\vb r)+n_\downarrow^2(\vb r) \right)
    +2g_{12}n_\uparrow(\vb r) n_\downarrow(\vb r)\right].
    \label{eq-Hinteraction}
\end{eqnarray}
Then we expand the wave function in terms of Bloch basis $\phi_{n, {\bf k}}$. Note that the Bloch wave vector ${\bf k}$ is well defined and conserved 
during scattering of Bloch states, similar to that in free space. The condensate should occur at one or both of the single-particle minima $\vb k=\pm \vb k_0$. 
By only considering interaction at these two vectors, we obtain an effective interaction over the Bloch bands,
\begin{eqnarray}
	\mathcal{V}_\mathrm{I}= U_{\vb k_0}\left( n_{\vb k_0}+n_{-\vb k_0} \right)^2+\left( V_{\vb k_0}-2U_{\vb k_0} \right)n_{\vb k_0}n_{-\vb k_0}.
    \label{eq-interaction2}
\end{eqnarray}
Here the two coefficients $g$ and $g_{12}$ will contribute to both $U_{\vb k_0}$ and $V_{\vb k_0}$ in a linear but complicated way. 
In general, $U_{\vb k_0} > 0$, thus the ground state of the condensate over these two degenerate points is purely determined by the sign of the second 
term \cite{zheng_collective_2012,chen_quantum_2018,li_quantum_2012,ji_experimental_2014}. It occupies a single vector (a plane wave phase BWP) when 
$V_{\vb k_0}-2U_{\vb k_0}> 0$ and two vectors (a stripe phase BST) with equal population when $V_{\vb k_0}-2U_{\vb k_0}< 0$. With this criterion, we determine the phase 
boundary between these two phases in Fig.~\ref{fig-fig3}(b). Strikingly, we find that the spin polarizations can fundamentally influence the scatterings in the 
condensate, thus dramatically influence the phase boundaries between these phases. The change of this boundary is further confirmed by numerical simulation using 
Gross-Pitaevskii equation (see Fig.~\ref{fig-fig3}(c) - (d)), in which during the transition from BPW phase to BST phase, dramatic changes in ground state
energy $E_g - E_g(g_{12}^c)$ and total spin polarization can be observed. 

In the special condition with only one Raman coupling, say $\Omega_2 = 0$. We find $U_{\vb k_0} = g-\eta g+\eta g_{12}$
and $V_{\vb k_0}-2U_{\vb k_0} = (6\eta-2)g+(2-2\eta)g_{12}$, where $\eta={2\Omega_1^2}/{k_1^4}$. The phase boundary is \cite{li_quantum_2012,
li_superstripes_2013}
\begin{eqnarray}
    \frac{g_{12}^c}{g}=\frac{2-6\eta}{2-2\eta}
    &=& \frac{k_1^4-6\Omega_1^2}{k_1^4-2\Omega_1^2},
    \qq{for}
    \abs{\Omega_1}<k_1^2/2,
    \label{eq-1dboundary}
\end{eqnarray}
which corresponds to the blue solid line in Fig. \ref{fig-fig3} (b). This boundary can be used to explain the four limits that when $\Omega_1 = 0$, 
$\Omega_2^\text{c} = \pm 1$, and  when $\Omega_2 = 0$, $\Omega_1^\text{c} = \pm 1$. 

\begin{figure}[h]
    \centering
    \includegraphics[width=0.48\textwidth]{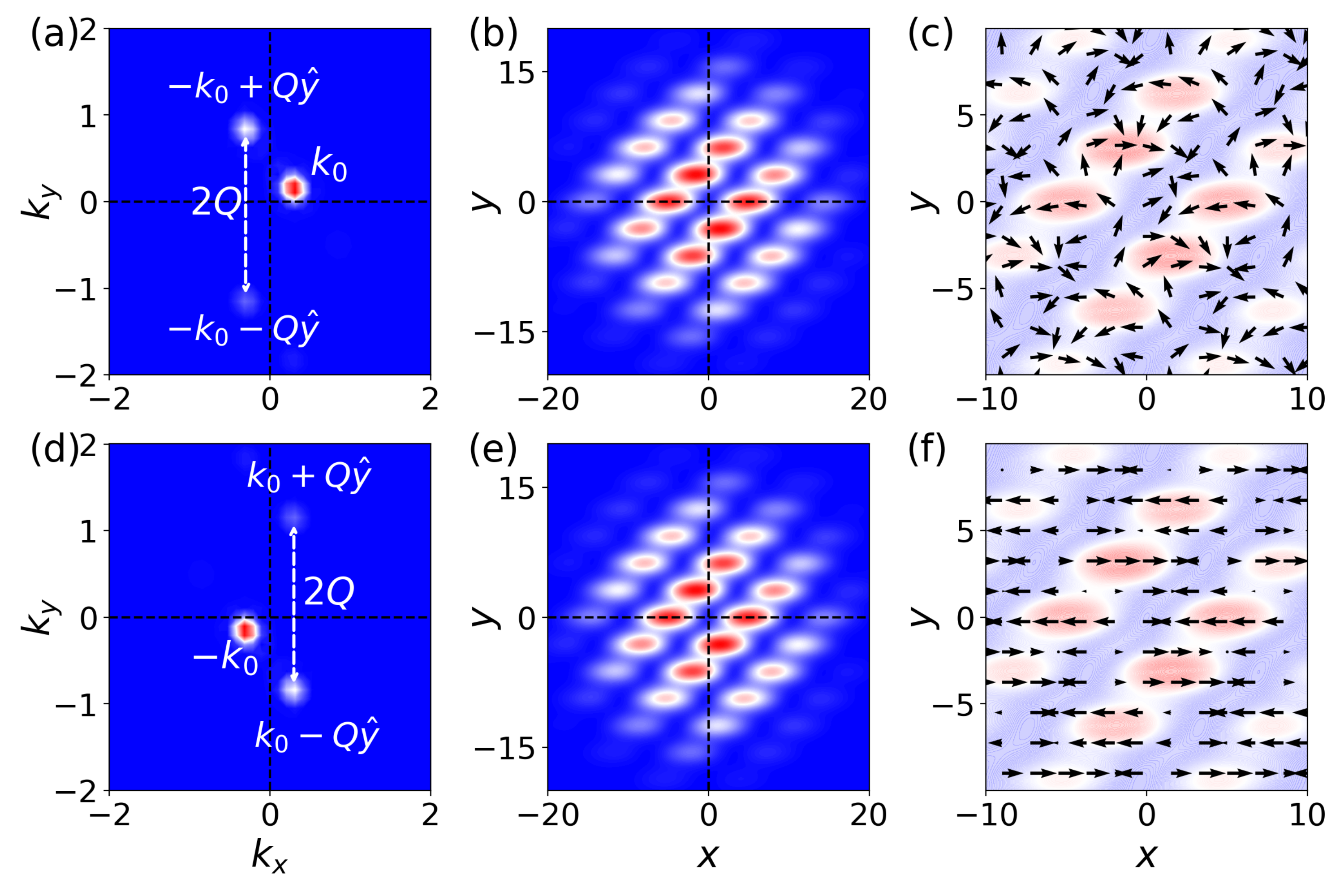}
	\caption{Properties of the BST phase with parameters $\Omega_1 = 0.4$, $\Omega_2 = 0.8$, $g_{12}/g = - 0.1$, $\vb k_0\approx(0.30,0.15)$.  
	Other descriptions are the same as that in Fig. \ref{fig-fig4}.}
	\label{fig-fig5}
\end{figure}

{\it Spin textures, density modulations and effective Hamiltonian}. 
We next discuss the properties of these two phases in Fig. \ref{fig-fig4} and Fig. \ref{fig-fig5}, which exhibit rich spin textures and density modulations 
in real space. In the BPW phase, only one Bloch wave vector ${\bf k}_0$ is occupied. Correspondingly, one spin component will occupy one plane wave momentum ${\bf k}_0$, and the other spin component occupy
two momenta ${\bf k}_0 \pm {\bf Q}$. The other momenta such as ${\bf k}_0 + n {\bf Q}$ for $n \ne \pm 1$ are presented, but not 
discernible in the present plot. By keeping only these three components as $(\phi_{\downarrow,{\bf k}_0,-\vb Q},\phi_{\uparrow,{\bf k}_0,0}, \phi_{\downarrow,{\bf k}_0,\vb Q})^T$, 
we obtain a minimal three-band effective Hamiltonian,
\begin{equation}
	\mathcal{H}_{\text{eff}} ({\bf k}_0) = \begin{pmatrix}
            \epsilon_{{\bf k_0-{\bf Q}}}^{-1}  & \Omega_1  & 0 \\
		\Omega_1^*  & \epsilon_{{\bf k}_0 }^{0}  & \Omega_2 \\
			0	& \Omega_2^*  & \epsilon_{{\bf k}_0 + {\bf Q}}^{+1}
	\end{pmatrix},
	\label{eq-H0}
\end{equation}
where $\epsilon_{{\bf k}}^{0} = ({\bf k}_{x} + K/2)^2 + {\bf k}_{y}^2$ and $\epsilon_{{\bf k}}^{\pm 1} = ({\bf k}_{x} - K/2)^2 + {\bf k}_{y}^2$. 
For this reason, the condensate should occupy multiple momenta simultaneously. It should be pointed out that the similar Hamiltonian in three hyperfine levels
has also been derived in Ref. \cite{lin2009boseeinstein} for the creation of light-induced vector gauge potential. We can use this effective Hamiltonian to 
understand the results in these two figures. 
In Fig. \ref{fig-fig4}, the two Raman couplings have different strengths, thus the two peaks in Fig. \ref{fig-fig4} (a) have different intensities. 
We find the ground state wave function to be $(-0.154,0.897,-0.413)^T$, thus the intensity ratio between these two peaks ${\bf k}_0 \pm {\bf Q}$ is $0.413/0.154=2.69$, 
while the GPE gives $2.58$. The intensity ratio between ${\bf k}_0$ and ${\bf k}_0 + {\bf Q}$ is $0.897/0.413\approx 2.17$, while the GPE gives $2.24$. 
The interference between different momenta can give rise to density modulation in real space (see Fig. \ref{fig-fig4} (b) and (e)), 
while in free space, this kind of modulation is absent. Furthermore, this phase will also exhibit some interesting spin textures in real space as shown in Fig. \ref{fig-fig4} (c) and (f). We then
compare these features to that in BST phase in Fig. \ref{fig-fig5}, in which each component will exhibit three peaks in momentum space due to occupation of both wave 
vectors $\pm {\bf k}_0$. As a result, the density modulations and spin textures in real space are also totally different. Note that in the BST phase, the total 
spin is balanced, thus $\langle \psi_{{\bf k}_0}|\sigma_z | \psi_{{\bf k}_0} \rangle = 0$ (see Fig. \ref{fig-fig5} (f)). These features can be understood from two 
copies of model (\ref{eq-H0}), i.e., $\text{diag}(\mathcal{H}_{\text{eff}} ({\bf k}_0), \mathcal{H}_{\text{eff}} (-{\bf k}_0))$. Due to the same parameters 
$\Omega_1$ and $\Omega_2$ used in both figures, they should have the same intensity ratios. These features can be used as smoking gun evidences in experiments to 
identify these two phases.

{\it Conclusion and discussion}. The model studied here possesses some features similar to that of two dimensional SOC, although it can never been reduced to the well-known 
Rashba or Dresselhaus SOC. In our model the TR symmetry and the real representation of $\mathcal{H}_{\bf k}$ ensure that the geometry phase around any closed 
loop \cite{pancharatnam_generalized_1956, longuet-higgins_studies_1958,berry_quantal_1984} 
exactly vanish. However, by applying an in-plane Zeeman field $h_x\sigma_x$, which still respects $\Theta_\text{tr}$ symmetry but breaks the 
$\mathcal{K}$ symmetry, we can awake the two dimensional nature of SOC with a finite geometry phase.
While this feature is not essential for BEC in Bloch bands, it may be important for the realization of topological superfluids in degenerate Fermi gases, in which the 
BCS pairing \cite{cooper_bound_1956,bardeen_microscopic_1957,wu_unconventional_2013,qu_fulde-ferrell-larkin-ovchinnikov_2014} is ensured by $\Theta_\text{tr}$ symmetry. 

To conclude we demonstrate some exotic condensates in the Bloch bands with ODPP realized using three laser beams coupled to the same three level system. 
By involving more lasers from the same source, different forms of ODPP can be realized. This kind of potential can never approach the tight-binding limit, 
thus it enables us to simulate some intriguing physics beyond the realm of condensed matter physics, which should 
be an important goal in AMO physics. This ODPP can change the interactions over the Bloch bands, thus may lead to new physics, such as collective
oscillation and damping of condensate \cite{liu1997theoretical,giorgini1998damping, wu2018beliaev}, polaron physics \cite{cucchietti2006strong, hu2016bose, shchadilova2016quantum}, topological superfluids \cite{gong_bcs-bec_2011,huang2015large} and their quench dynamics.

\textit{Acknowledgements.} M.G. is supported by the National Youth Thousand Talents Program (No. KJ2030000001), the USTC start-up funding (No. KY2030000053), 
the national natural science foundation (NSFC) under grant No. 11774328 and National Key Research and Development Program of China (No. 2016YFA0301700). HP is supported by the US NSF.

\bibliography{ref}

\end{document}